\documentclass[useAMS,usenatbib,usegraphicx]{mn2e}
\newcommand\lsim{\mathrel{\rlap{\lower4pt\hbox{\hskip1pt$\sim$}}
    \raise1pt\hbox{$<$}}}
\newcommand\gsim{\mathrel{\rlap{\lower4pt\hbox{\hskip1pt$\sim$}}
    \raise1pt\hbox{$>$}}} \newcommand{\dm}{\mathrm {dm}}
\newcommand{\br}{\mathrm {b}} 
 \newcommand{\tot}{\mathrm{tot}}

\begin{document}
\title[Gas rich halos - simulation vs.\ linear theory]{The nonlinear 
evolution of baryonic overdensities
in the early universe: Initial conditions of numerical simulations}

\author[Smadar Naoz, Naoki Yoshida \&  Rennan Barkana ]
{Smadar Naoz $^{1,2}$\thanks{E-mail: snaoz@northwestern.edu } ,
Naoki Yoshida$^3$ \& Rennan Barkana$^2$\\
$^{1}$ CIERA, Northwestern University, Evanston, IL 60208,
USA\\
$^{2}$ Raymond and Beverly Sackler School of Physics and
Astronomy, Tel Aviv
University, Tel Aviv 69978, Israel \\
$^{3}$ IPMU, University of Tokyo, 5-1-5 Kashiwanoha, Kashiwa, Chiba 277-8583, Japan}  \maketitle

\begin{abstract}

We run very large cosmological $N$-body hydrodynamical simulations in
order to study statistically the baryon fractions in early dark matter
halos. We critically examine how differences in the initial conditions
affect the gas fraction in the redshift range $z=11-21$. We test three
different linear power spectra for the initial conditions: (1) A
complete heating model, which is our fiducial model; this model
follows the evolution of overdensities correctly, according to
\citet{NB05}, in particular including the spatial variation of the
speed of sound of the gas due to Compton heating from the CMB.  (2) An
equal-$\delta$ model, which assumes that the initial baryon
fluctuations are equal to those of the dark matter, while conserving
$\sigma_8$ of the total matter.  (3) A mean $c_s$ model, which assumes
a uniform speed of sound of the gas.  The latter two models are often
used in the literature. We calculate the baryon fractions
for a large sample of halos in our simulations. Our fiducial model
implies that before reionization and significant stellar heating took
place, the minimum mass needed for a minihalo to keep most of its
baryons throughout its formation was $\sim 3 \times
10^4$~M$_\odot$. However, the alternative models yield a wrong (higher
by about $50\%$) minimum mass, since the system retains a memory of
the initial conditions. We also demonstrate this using the "filtering
mass" from linear theory, which accurately describes the evolution of
the baryon fraction throughout the simulated redshift range.

\end{abstract}

\section{Introduction}\label{intro}

Recent measurements of anisotropies of the cosmic microwave background
(CMB) radiation have revealed the detailed distribution of matter in
the Universe a few hundred thousand years after the Big Bang \citep{Spergel+03,Spergel+07,wmap5,Komatsu+10}.  Observations
utilizing large ground-based telescopes and space telescopes have
discovered galaxies and black holes that were in place when the age of
the Universe was less than a billion years.  Moreover, many galaxies
have been found at $z > 7$ \citep{hubble8,hubble6-9} in the Hubble
Ultra Deep Field, whereas already a few gamma-ray bursts at $z > 6$
have been detected by the {\em Swift} satellite
\citep[][]{GRB8a,GRB8natur,lin09}.  These first objects are probably
the building blocks of the present day galaxies, thus, solving the
puzzle behind their formation will have a profound implication on our
understanding of the Universe \citep[see for recent reviews][and
references therein]{Bromm+09,Y09_early}.
%

The formation of the first generation of galaxies in the Universe has
been studied for many years. High resolution cosmological simulations
can follow complex astrophysical processes, while analytical
calculations can provide an over-all understanding, and can be used to
decouple different physical effects seen in simulations.  Analytic
models are also useful for estimating the limitations of numerical
simulations such as insufficient resolution and small boxsizes
(Yoshida et al. 2003; Barkana \& Loeb 2004; Naoz \& Barkana 2005).
Combining the two approaches may offer many of the advantages of both.

The initial conditions (hereafter ICs) in a cosmological simulation
can have a large effect on the formation of the first galaxies in
simulations, i.e., both on the formation time (or on the halo
abundance at a given time) and the halo properties at formation time (such
as the average gas fraction). \citet{Y03initial} studied high-redshift
structure formation and reionization while testing two different
models for power spectra as their ICs. They found that different
models have a profound effect on the abundance of primordial
star-forming gas clouds and thus on when the reionization was
initiated and its progress. In the analytical point of view,
\citet{NNB} and \citet{NB07} showed that the ICs at high redshift have
a significant effect on the halo abundance and the gas fraction at
virialization. While these effects are largest at the highest
redshift, e.g., $z \sim 65$ for the first star in the universe
\citep{NNB}, they are still significant for halos forming at $z\sim 10
- 30$. The first gas rich halos at these redshifts are expected to
host the first stars \citep[$z\sim
65-30$][]{NNB,Yoshida06,Gao07,Trenti_z} and even the first gamma-ray
bursts \citep[e.g.,][]{BrL06,NBr07}. Thus, investigating the formation
properties of these halos is of prime importance .

Gas rich halos in the early Universe may very well be a nurturing
ground for dwarf galaxies, which at high redshift can form stars
\citep[e.g.,][and references
therein]{BCL02,BCL99,Abel+02,Yoshida+06,Y08_firstS} perhaps even at a
high star formation rate
\citep{Ricotti+02,Greif10,Clark10}. Their properties are very
important as they are responsible for metal pollution and the ionizing
radiation at these early times
\citep[e.g.,][]{Shapiro+04,Ciardi+06,Gnedin+08,Trenti+09}.  Moreover,
halos that are too small for efficient cooling via atomic hydrogen,
i.e., minihalos, are most susceptible to the effect of initial
conditions.  While they may not normally host astrophysical sources,
minihalos may produce a 21-cm signature (\citet{Kuhlen,Shapiro06,NB08}
but see \citet{Furlanetto06}), and they can block ionizing radiation
and produce an overall delay in the initial progress of reionization
\citep[e.g.,][]{bl02, iliev2, iss05, mcquinn07}.  The evolution of the
halo gas fraction at various epochs of the universe is of prime
importance, particularly in the early universe.

In this paper, we examine the effect of using different initial
conditions in simulations on the resulting minimum gas-rich halo mass
in the redshift regime $z=11-21$.  We perform Gadget-2
\citep{Gadget,G2} simulations using a total of $768^3\times 2$
particles.  We compare the initial conditions presented in
\citet{NB05}, which describe the linear evolution of overdensities in
a fully consistent way, to two other alternative ICs, often used in
the literature. We also compare to the prediction of the gas-rich mass
from linear theory.  We describe our different initial conditions and
simulations in sections
\ref{ICs} and \ref{sim}, respectively. Our simulation results are
presented in section \ref{res} where we divide our discussion to the
evolution of the non-linear power spectra (section \ref{nonlin}) and
to the minimum gas-rich halo mass resulting from either linear theory
or from the simulations (section \ref{Mc_f}). Finally, we discuss our
conclusions (section \ref{dis}).

Throughout this paper, we adopt the following cosmological
parameters: ($\Omega_\Lambda$, $\Omega_{\rm M}$, $\Omega_b$, n,
$\sigma_8$, $H_0$)= (0.72, 0.28, 0.046, 1, 0.82, 70 km s$^{-1}$
Mpc$^{-1}$)  \citep{wmap5}.

\section{Different Initial Condition models - Basic Equations}\label{ICs}
\subsection{The fiducial ICs - "fid"}\label{correct}

We follow \citet{NB05}, who studied the linear evolution
of both dark matter and baryon overdensities.  
The fluctuations of the temperature of the baryons
($\delta_T$) cannot be described as a simple function of a
spatially uniform baryonic sound speed $c_s(t)$, as was previously
assumed \citep[e.g.,][]{Ma}. 
Furthermore, at high redshifts, 
the baryon density fluctuations ($\delta_\br$) are not equal to
those of dark matter ($\delta_{\dm}$) 
\citep[contrary to a common assumption in simulations;
four redshift examples are shown in figure 1 of ][]{NB05}. We label
the power spectrum model as the "fid" (fiducial) ICs since
it follows the evolution of linear overdensities in a complete
and consistent way.

\begin{figure}
  \centering \includegraphics[width=84mm]{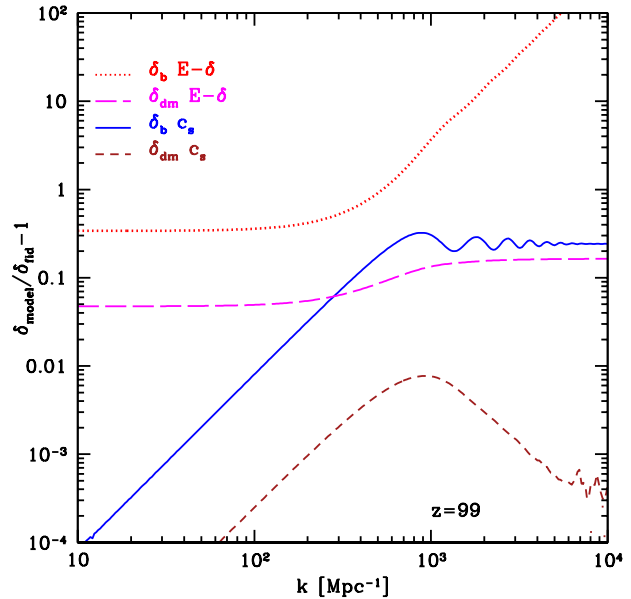}
\caption{The relative difference (specifically,
$\delta_{model}/\delta_{ch}-1$) between the fiducial linear initial
conditions and the alternative models at $z=99$. We consider the
relative difference between the fid ICs and the mean $c_s$ ICs for both
the baryons and dark matter (solid and short-dashed curves,
respectively), and the relative difference between the fid ICs and the
E-$\delta$ ICs for both the baryons and dark matter (dotted and
long-dashed curves, respectively). Note that we have plotted here the
absolute value; the mean $c_s$ model gave a negative value (i.e., an
underestimate compared to the fid model) while the E-$\delta$ model
gave a positive value (i.e., an overestimate).} \label{fig:ICs}
\end{figure}

Following \citet{NB05} we write the basic equations that
describe the evolution of the dark matter, baryon density and
temperature fluctuations:
\begin{equation}
 \ddot{\delta}_{\dm}
 + 2H \dot {\delta}_{\dm} = \frac{3}{2}H_0^2\frac{\Omega_m}{a^3}
\left(f_{\br} \delta_{\br} + f_{\dm} \delta_{\dm}\right) \ , \label{eq:dm}
\end{equation}
where $f_{\dm}$  and $f_{\br}$ are the mean cosmic dark matter and baryonic fraction respectively.
Here we follow the standard notations for cosmological 
parameters such as $\Omega_m, H_0$. 
The baryons are also subject to a pressure term:
\begin{equation} \label{eq:b}
\ddot{\delta}_{\br}+ 2H \dot {\delta}_{\br} =
\frac{3}{2}H_0^2\frac{\Omega_m}{a^3} \left(f_{\br} \delta_{\br} +
f_{\dm} \delta_{\dm}\right)-\frac{k^2}{a^2}\frac{k_B\bar{T}}{\mu}
\left(\delta_{\br}+\delta_{T}\right)\ ,
\end{equation}
where $\mu$ is the mean molecular weight, $k_B$ is the Boltzmann
constant and $k$ is the wavenumber.  Using the first law of
thermodynamics, \citet{NB05} derived the equations for the
evolution of the baryon average temperature and temperature
fluctuations:
\begin{equation}
\label{mean} \frac{d \bar{T}} {dt} = - 2 H \bar{T} +
\frac{x_e(t)}{t_\gamma}\, (\bar{T}_\gamma - \bar{T})\, a^{-4}\ ,
\end{equation}
where $\bar{T}_\gamma=[2.725\ {\rm K}]/a$ is the mean CMB temperature,
and the first-order equation for the perturbation:
\begin{equation}
\label{gamma} \frac{d \delta_T} {d t} = \frac{2}{3} \frac{d
\delta_\br} {dt} + \frac{x_e(t)} {t_\gamma}a^{-4} \left\{
\delta_\gamma\left( \frac{\bar{T}_\gamma}{\bar{T}} -1\right)
+\frac{\bar{T}_\gamma} {\bar{T}} \left(\delta_{T_\gamma} -\delta_T
\right) \right\}\ , 
\end{equation}
with the second term on the right-hand side accounting for the Compton
scattering of the CMB photons on the residual electrons from
recombination, where $x_e(t)$ is the electron fraction out of the
total number density of gas particles at time $t$, and
\begin{equation}
\label{tgamma} t_\gamma^{-1} \equiv \frac{8} {3} \bar{\rho}_\gamma^0
\frac{\sigma_{T}\, c} {m_e} = 8.55 \times 10^{-13} {\mathrm{yr}}^{-1}\ ,
\end{equation}
where $\sigma_{\rm T}$ is the Thomson scattering cross section
and $\rho_\gamma$ is the photon energy density. The first term on the
right-hand-side of each of these two equations (\ref{mean}) and
(\ref{gamma}) accounts for adiabatic expansion of the gas, and the
remaining terms capture the effect of the thermal exchange with the
CMB.  Following
\citet{NB05} we have numerically calculated the evolution of the
perturbations by modifying the CMBFAST code \citep{cmbf} according
to these equations. Note that similar physics was also explored by
\citet{Naoshi1,Naoshi2}.

We solve the complete set of equations to obtain the power spectrum at
different redshifts which can be used as initial conditions for our
simulations. Figure \ref{fig:ICs} shows the ratio between this initial
condition to the two alternative models tested in this paper.

\subsection{Alternative model I - equal $\delta$ - "E-$\delta$"}\label{sec:Edelta}

In many cosmological ICs for N-body simulations and semi-analytical
calculations, the fluctuations of the baryons are assumed to be equal
to the fluctuations of the dark matter. We construct a model that
includes this incorrect assumption while maintaining the correct
overall $\delta_{tot}$ (i.e., conserving $\sigma_8$ at $z=0$, see appendix \ref{sig8} for more details). Thus, in our
``E-$\delta$'' model we calculate the correct $\delta_{tot}$ as a
combination of $\delta_\br$ and $\delta_{\dm}$ from the fiducial
 calculation in section (\ref{correct}), but then take the
baryon perturbation to be the same as for the dark matter, namely:
\begin{equation}
\delta_\br^{E\delta}=\delta_\dm^{E\delta}=\delta_{tot}^{ch}=f_\br\delta_\br^{ch}+f_\dm\delta_\dm^{ch}  \ ,
\end{equation}
where $\delta_{\br,\dm}^{ch}$ ($\delta_{\br,\dm}^{E\delta}$) is the
resulting linear over-density from the fiducial calculation
(E-$\delta$ model) for the baryons and dark matter, respectively. We
then compare the equal $\delta$ model to our fiducial
calculation. Figure \ref{fig:ICs} shows the ratio between the fid ICs
and the E-$\delta$ model for both the baryons and dark matter.  We
find that the E-$\delta$ model overestimates the baryon fluctuations
by $\gsim 30 \%$ on large scales ($k^{-1} \gsim 10$~kpc) while the
overestimate grows to a much larger factor on small scales.

Before recombination the baryons were tightly coupled to the
radiation, resulting in suppression of the growth of their
overdensity. However, the dark matter component, which is not affected
by the photons, could basically grow once the fluctuation wavelength
entered the Hubble horizon (in the linear regime, before equality, the
dark mater fluctuations grew logarithmically with the scale factor,
where after equality they grew linearly with the scale
factor). Therefore, this resulted in a suppression of the baryonic
overdensity by about three orders of magnitude compare to the dark
matter at recombination
\citep[e.g., fig.~1 in][]{NB05}. While the baryons subsequently 
fall into the potential wells of the dark matter, it takes them some
time to catch up, and the baryon fluctuations are still suppressed
even at lower redshifts. This point is often overlooked in simulations
and analytical calculations.

\subsection{Alternative model II - the mean
  sound speed approximation - "mean cs"}\label{sec:cs}

\citet{NB05} showed that the presence of spatial fluctuations in the
sound speed modifies the calculation of perturbation growth
significantly. Nevertheless, for completeness and as a case of comparison
with previous results, we compare the simulation results with the
results obtained using this approximation. 
Thus, we proceed by presenting the basic equations of the growth of
density fluctuations, in this approximation of a uniform sound speed
(hereafter ``mean $c_s$''). The evolution of the density fluctuations
is described by a different set of coupled second order differential
equations:
\begin{eqnarray}
\label{g_cs}
 \ddot{\delta}_{\dm}
 + 2H \dot {\delta}_{\dm} & = & \frac{3}{2}H_0^2\frac{\Omega_{m}}{a^3}
\left(f_{\br} \delta_{\br} + f_{\dm} \delta_{\dm}\right)\ , \\ 
\ddot{\delta}_{\br}+ 2H \dot {\delta}_{\br} & = &
\frac{3}{2}H_0^2\frac{\Omega_{m}}{a^3} \left(f_{\br} \delta_{\br} +
f_{\dm}
\delta_{\dm}\right)-\frac{k^2}{a^2}c^2_s\delta_{\br}\ ,\nonumber
\end{eqnarray}
where $c_s^2=dp/d\rho$ is assumed to be spatially uniform (i.e.,
independent of $k$) and is thus calculated from the thermal evolution
of a uniform gas undergoing Hubble expansion. With this assumption,
the temperature fluctuations (as a function of $k$) are simply
proportional at any given time to the gas density fluctuations:
\begin{equation}
\frac{\delta_T}{\delta_{\br}}=\frac{c_s^2}{k_B \bar{T}/\mu}-1\ .
\end{equation}
\citet{NB05} showed that this approximation leads to an underestimation 
of the baryon density fluctuations by up to 30\% at $z=100$ and 10\%
at $z=20$ for large wavenumbers. Figure~\ref{fig:ICs} shows the ratio
between the mean $c_s$ initial conditions and the fiducial
ones for both the baryons and dark matter.  It agrees with our
previous results, showing that the underestimate by the mean $c_s$
model is greatest at $k^{-1} \sim 1$~kpc. The non-linear evolution
resulting from these initial conditions will result in shallower
potential wells compared to the fiducial calculation,

Even though it is clear that the precise baryon temperature
fluctuations at high redshift are very significant, still many
simulations use initial conditions that assume a uniform speed of
sound in the Universe. As shown below this leads to significantly
different estimates for the gas content of the early halos.

\section{The simulation}\label{sim}

\subsection{Basic parameters}\label{sec:basic}

We run a Gadget 2 simulation \citep{Gadget,G2} starting from redshift
$99$, for a total of $2\times768^3$ particles ($768^3$ particles each
for the Dark Matter and baryon components) and our box size is:
$2$~Mpc. We choose this box size so that a halo mass of
$10^5$~M$_{\odot}$ would have $\sim500$ particles. This way according
to \citet{NBM} we are able to estimate the gas fraction in
$\sim10^5$~M$_{\odot}$ halos correctly (see below for the halo
definition).  Our softening length is 0.2 comoving kpc.

For all runs, glass-like cosmological ICs were generated using the
Zel'dovich approximation. The transfer functions were generated using
the various models described above.
We have used a glass file which was randomly displaced thus removing
the coupling between nearby DM and gas particles. Using this
randomization procedure we achieve essentially the same effect to that
shown in \citet{Yoshida03b}. In generating the ICs,
a convolution between the glass file and the transfer function from
the different models was done, thus taking into account the different
velocities of the DM and baryons (for the fiducial and mean cs
models). We note that we have used the same phases for the DM and
baryons, in all of the simulations.


We set the initial temperature to be $164.11$~K (as derived from
linear theory), and thus Gadget assumes neutral and monoatomic gas,
and converts to thermal energy (i.e., adiabatic initial conditions). Although this work emphasizes the need
for a precise calculation of the baryon overdensities resulting from
temperature fluctuations, we actually neglect the temperature
fluctuations in the initial conditions. This may not be a bad
approximation since the halos we study are already somewhat non-linear
at our initial redshift, and the Compton heating is quite small
compared to the adiabatic heating during non-linear gravitational
collapse (see Appendix \ref{Cheating}). A more complete treatment would
be to include in the simulation the precise temperature fluctuations,
which we leave for future work. Nevertheless, even with the current
treatment our results show consistency with linear theory.

\subsection{Halo definition}

We locate dark matter halos by running a FOF group-finder algorithm
with a linking parameter of $0.2$.  We then find the center of mass of
each halo and calculate the density profile of the dark matter and
baryons, separately. In order to derive the density profile we assume
a spherical halo, and divide it to 2000 shells.  Combining these
density profiles, we find the virial radius $r_{vir}$ at which the
overdensity is $200$ times the background density, and the
gas fraction of each halo.

Recently, \citet{Trenti_halos} performed a resolution analysis in
order to study the mass definition of halos in simulations. Their
conclusion (their figure 2) is that using the FOF algorithm and
assuming about 500 particles per spherical halo introduces an error of
$\sim 15\%$ in the mass definition.  In our gas fraction analysis we
have chosen only halos with a number of particles larger or equal to
$500$, i.e., we limit our errors in halo mass definition to below
$\sim 15 \%$. Also, according to \citet{NBM}, this way we can estimate
the gas fraction inside a halo accurately.

\section{Results and comparison among the models}\label{res}

\subsection{Non-linear power spectrum evolution}\label{nonlin}

One way to probe cosmic structure particularly on small scales is
through the non-linear power spectrum. We begin our simulation at
$z=99$ with linear initial conditions\footnote{This is, of course, an
approximation, ,since as shown in \citet{NNB} at $z=99$ overdensities
are already slightly non-linear. The effect of starting the simulation
at high redshifts is studied elsewhere (Naoz et.\ al, in prep.)}.  The
main disagreement between the three models lies in the baryonic
component (although the E-$\delta$ calculation also underestimates the
dark matter overdensities by $\sim 10 \%$ ). This input difference is
then modified by the non-linear evolution. 

%

\begin{figure}
  \centering \includegraphics[width=84mm]{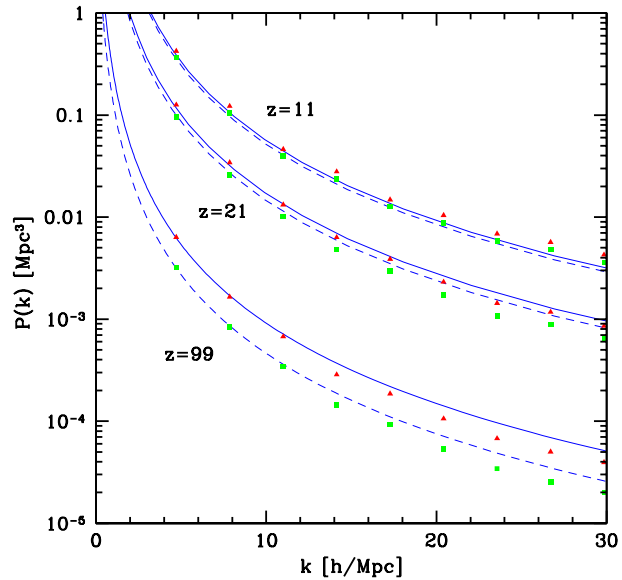}
\caption{Comparison of the linear and non-linear power spectra. The linear
power spectra \citep[generated according to ][]{NB05} are shown
for the dark matter and baryon components (solid and dashed curves,
respectively), while the corresponding non-linear spectra (as measured
in the simulations) are shown as triangles and squares, respectively.
We show results for the fid model at redshifts $z = 99$, $21$ and $11$.    } \label{fig:lin_non}
\end{figure}

Following \citet{Yoshida03b} we compared the
linear power spectrum for the fid model, as computed from \citet{NB05}, for the dark matter and baryon components, with the
non-linear power spectra from the simulation (see figure \ref{fig:lin_non}). The two
power spectra agree well as expected in the linear regime. We note
that the other two models approach the fid model at low redshifts (see
appendix \ref{sig8} figure \ref{fig:PSev}).
%

Fig \ref{fig:nonlinear} shows the differences among the fid, E-$\delta$
and mean $c_s$ ICs, in terms of the non-linear power spectra at the
later redshifts at which halos were formed in our simulation. The mean
$c_s$ model maintains over time roughly the same level of discrepancy
with the fid model, while in the E-$\delta$ model both the baryonic and
dark matter differences decline slightly slower than with the inverse
scale factor. As clearly can be seen from figure~\ref{fig:nonlinear},
the non-linear evolution of halos is still strongly affected by the
choice of initial conditions even at redshift $12$. The fid ICs
\citep{NB05} describe the linear evolution consistently
and thus represent the best available prescription for the initial
conditions.

\begin{figure}
  \centering \includegraphics[width=84mm]{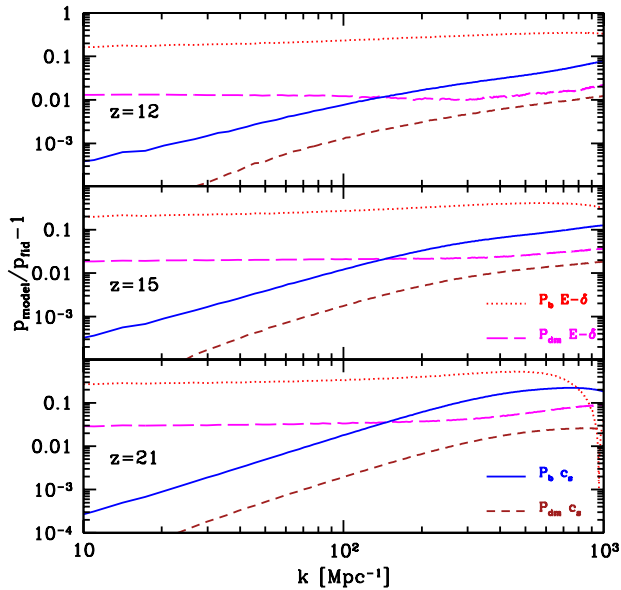}
\caption{The ratio of the non-linear power spectra (specifically,
$P_{\rm model}/P_{\rm fid}-1$) at $z=21$, 15, and 12 (from bottom to top);
curves are denoted as in figure~\ref{fig:ICs}. Note that we have
plotted here the absolute value; the mean $c_s$ model underestimates
and the E-$\delta$ model overestimates the power spectrum compared to
the fid model. } \label{fig:nonlinear}
\end{figure}

\subsection{The minimum gas rich mass }\label{Mc_f}

Studying the galaxy evolution and reionization either by using
simulations (both AMR and SPH) or by using analytical calculations
relies on knowing the amount of gas within the dark matter halos. The
simplest assumption, often used in the literature, is that a dark
matter halo has the mean cosmic fraction. This can lead to incorrect
results, especially when one tries to study star formation, galaxy
mergers, and related phenomena.

Consider the various scales involved in the formation of
non-linear objects containing DM and gas. On large scales (small
wavenumbers) gravity dominates halo formation and gas pressure can
be neglected. On small scales, on the other hand, the pressure
dominates gravity and prevents baryon density fluctuations from
growing together with the dark matter fluctuations. The relative
force balance at a given time can be characterized by the
\citet{jeans} scale, which is the minimum scale on which a small
gas perturbation will grow due to gravity overcoming the pressure
gradient. As long as the Compton scattering of the CMB on the residual free electrons after cosmic
recombination kept the gas temperature coupled to that of the CMB, the
Jeans mass was constant in time. However, at $z\sim 200$ the gas
temperature decoupled from the CMB temperature and the Jeans mass
began to decrease with time as the gas cooled adiabatically. Any
overdensity on a scale more massive than the Jeans mass at a given
time can begin to collapse, due to a lack of sufficient
pressure. However, the Jeans mass is related only to the evolution of
perturbations at a given time. When the Jeans mass itself varies with
time, the overall suppression of the growth of perturbations depends
on a time-averaged Jeans mass.

\citet{cs} defined a ``filtering mass'' that describes the highest
mass scale on which the baryonic pressure still manages to
suppress the linear baryonic fluctuations significantly.
\citet{gnedin00} suggested, based on a simulation, that the
filtering mass also describes the largest halo mass whose gas content
is significantly suppressed compared to the cosmic baryon
fraction. The latter mass scale, in general termed the
``characteristic mass'', is defined as the halo mass for which the
enclosed baryon fraction equals half the mean cosmic fraction.  Thus,
the characteristic mass distinguishes between gas-rich and gas-poor
halos. Many semi-analytical models of dwarfs galaxies often use the
characteristic mass scale in order to estimate the gas fraction in
halos
\citep[e.g.,][]{Bullock,Benson02a,Benson02b,Somerville}. Theoretically
this sets an approximate minimum value on the mass that can still form
stars.

\subsubsection{Prediction from linear theory}\label{sec:Mf}

In linear theory the filtering mass, first defined by \citet{cs},
describes the highest mass scale on which the baryon density
fluctuations are suppressed significantly compared to the dark
matter fluctuations.  \citet{NB07}  included the fact that the
baryons have smoother ICs than the dark matter
\citep[see][]{NB05} and found a lower value of the filtering mass
(by a factor of $3-10$, depending on the redshift). Following
\citet{NB07}, the filtering scale (specifically, the filtering
wavenumber $k_F$) is defined by expanding the ratio of baryonic to
total density fluctuations to first order in $k^2$:
\begin{equation}
\frac{\delta_\br}{\delta_\tot}=1-\frac{k^2}{k_F^2}+r_{\rm LSS}\ ,
\label{kf_btot}
\end{equation}
where $k$ is the wavenumber, and $\delta_\br$ and $\delta_\tot$ are
the baryonic and total (i.e., including both baryons and dark matter)
density fluctuations, respectively. The parameter $r_{\rm LSS}$ (a
negative quantity) describes the relative difference between
$\delta_\br$ and $\delta_\tot$ on {\em large scales}
\citep{NB07}, i.e.,
\begin{equation}
r_{\rm LSS} \equiv \frac{\Delta}{\delta_\tot}\ ,
\label{r_LSS}
\end{equation}
where $\Delta=\delta_\br-\delta_\mathrm{tot}$, \citep[see
also][]{BL05}.  The ratio $r_{\rm LSS}$ is independent of $k$, and its
magnitude decreases with time approximately $\propto 1/a$, since
$\Delta$ is roughly constant and $\delta_\tot$ is dominated by the
growing mode $\propto a$ \citep[see figure 1 top panel in][]{NB07}.

The filtering mass is
defined from $k_F$ simply as:
\begin{equation}
M_F=\frac{4\pi}{3}\bar{\rho_0}\left(\frac{1
}{2}\frac{2\pi}{k_F}\right)^3\ , \label{Mf}
\end{equation}
where $\bar{\rho_0}$ is the mean matter density today. This relation
is one eighth of the definition in \citet{gnedin00} (who also used a
non-standard definition of the Jeans mass).  In figure~\ref{fig:Mc}
(bottom panel) we show the filtering mass (solid curve) resulting from
eq.~(\ref{Mf}), as calculated in \citet{NB07} (see also their
figure~3).

For each of the models we calculate the filtering mass as described
here, assuming the model's initial conditions. Since the simulation is
limited in box size, all of the perturbations on large scales are
effectively frozen in the simulation. Therefore, we do not extract
$r_{\rm LSS}$ directly from the simulations, but instead calculate it
based on the initial conditions as $r_{\rm LSS} =
\Delta_{in}/(\delta_{\tot,in} a)$, where the subscript "in" refers to
initial. Thus, for example, for the E-$\delta$ case, $r_{\rm LSS}=0$.
Figure~\ref{fig:Mc} (bottom panel) shows
the analytical results of the filtering mass for the fid calculation,
the mean $c_s$ approximation and E-$\delta$ (solid, dashed and dotted
curves, respectively). Since the fid calculation is the most
consistent calculation, we compare the two other models to it.

The filtering mass represents the competition between gravity and
pressure, as it measures the largest scale at which pressure has had a
significant overall effect on halo formation. Since it measures an
integrated effect over the formation, this mass scale is also very
sensitive to the evolution history and the initial conditions
\citep[as shown in][]{NB07}. In the mean $c_s$ model, the temperature
fluctuations are greatly overestimated on all relevant scales
\citep[see][]{NB05}, while in reality the coupling to the CMB
(in the fid model) keeps the temperature fluctuations highly suppressed
for some time after recombination.
Moreover, as mentioned in section \ref{sec:basic} (and see also Appendix \ref{Cheating}), 
we do not include explicitly the effect of initial
temperature fluctuations in the simulations. However, the temperature
fluctuations from higher redshifts influence the baryon density at the
initial redshift
(see figure \ref{fig:ICs}) and suppress the baryon density on
small scales. As demonstrated in \citet{NB07}  the system
remembers the initial conditions. In other words, the initially
enhanced filtering mass (compared to the fid model) helps maintain a
higher filtering mass even at moderately low redshift.

In the E-$\delta$ model, the baryon perturbations start out
much higher than in the other models, so one might expect that the
final baryon fraction in halos would tend to be higher as well; here,
however, it is important to separate two issues. The high initial
baryon perturbations in the E-$\delta$ model are present at all
scales, so they affect even high-mass halos that are unaffected by
pressure. This can explain why the simulation with the E-$\delta$ ICs
produced the highest baryon fraction in high-mass halos (see the top
panel of Figure~\ref{fig:Mc}). However, when we consider the
differences between large and small scales, the high baryon
perturbations produce a large pressure term, increasing the effect of
pressure relative to gravity and producing a higher filtering mass in
the E-$\delta$ model than in the fid model. Note that the filtering
mass is particularly sensitive to the importance of pressure at the
very highest redshifts (above 100), since at lower redshifts the gas
cools and the Jeans mass decreases, reducing the contribution of these
redshifts to the final filtering mass.

%

We note that in \citet{NB07} the calculation of the filtering mass in
the fiducial model was compared to the time integrated
filtering mass in a model that assumes the mean speed of sound model,
neglects the $r_{\rm LSS}$ factor, and starts out with initial
conditions as in the E-$\delta$ model. Here, we have separated our
discussion into several different cases.


\subsubsection{The non-linear characteristic mass }\label{sec:Mc}
There is no apriori reason to think that the filtering mass can also
accurately describe properties of highly non-linear, virialized
objects. For halos, \citet{gnedin00} defined a characteristic mass
$M_c$ for which a halo contains half the mean cosmic baryon fraction
$f_b$. In his simulation he found the mean gas fraction in halos of a
given total mass $M$, and fitted the simulation results to the
following formula:
\begin{equation}
\label{f_g-alpha}
f_{g,\rm calc}= f_{\br,0} \bigg[1+\left(2^{\alpha/3}-1
\right)\left(\frac{M_c}{M}\right)^\alpha \bigg]^{-3/\alpha} \ ,
\end{equation}
where $f_{\br,0}$ is the gas fraction in the high-mass limit.  In this
function, a higher $\alpha$ causes a sharper transition between the
high-mass (constant $f_g$) limit and the low-mass limit (assumed to be
$f_g \propto M^3$). \citet{gnedin00} found a good fit for $\alpha =
1$, with a characteristic mass that in fact equaled the filtering mass
by his definition. By our definition, the claim from
\citet{gnedin00} is that $M_c=8\times M_F$.

\citet{NBM} found that, given their errors, the
filtering mass from linear theory is consistent with the
characteristic mass fitted from the simulations, for two
(pre-reionization) scenarios that they tested: the NoUV case (i.e., no
stellar heating) and the Flash case (i.e., after a sudden flash of
stellar heating). For clarity, we emphasize that this statement
($M_c=M_F$) refers to our definition of $M_F$ in equation~(\ref{Mf}).

The characteristic mass is essentially a non-linear version of the
filtering mass, and so it also measures the competition between
gravity and pressure. At high masses, where pressure is unimportant,
$f_g\to f_{b,0}$, while the low mass tail is determined by the
suppression of gas accretion caused by high baryonic pressure. 

\subsubsection{Comparison between the simulation and the 
theoretical predictions}

\begin{figure}
  \centering \includegraphics[width=84mm]{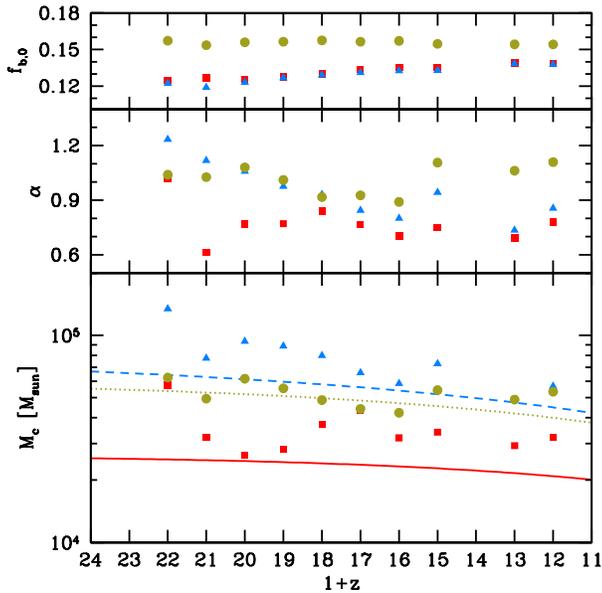}
\caption{ The parameters of the best fits in the form of 
equation~\ref{f_g-alpha}; different panels show $M_c$, $\alpha$, and
$f_{\br,0}$. We consider the fiducial calculation, mean $c_s$
approximation and the E-$\delta$ model (boxes, triangles and circles,
respectively), where we fit equation~(\ref{f_g-alpha}) to all data
points from halos with at least 500 particles. In the bottom panel we
also show the analytical calculation following
\citet{NB07}, for all the models, assuming the same
ICs as in the simulations (solid, dashed and dotted curves for fid,
mean $c_s$ and E-$\delta$, respectively).  We note that at $1+z=13$
the mean $c_s$ and the E-$\delta$ models have the same value of $M_c$, and that the fid model and the E-$\delta$ overlap at $1+z=17$.
We also note that the data for $1+z=14$ was unavailable due to a
computer failure.}
\label{fig:Mc}
\end{figure}

A major conclusion of the simulation results is that different ICs
result in different gas fractions in the final halos. Specifically, we
measure these differences through the characteristic mass at various
redshifts.  varies for different ICs.  We determine for each
simulation output the characteristic mass and the parameter $\alpha$
using a two-dimensional fit to equation~(\ref{f_g-alpha}), with
$f_{b,0}$ separately fixed to equal the average of the highest few
mass bins (see Appendix \ref{app} for a complete description of the fitting
process, together with the $1-\sigma$ errors).  In figure
\ref{fig:Mc} we show $f_{b,0}$, $\alpha$ and $M_c$, for all
the simulated cases. The characteristic mass clearly depends on the
initial conditions, with the mean $c_s$ model and E-$\delta$ model
both yielding gas suppression at systematically higher halo masses
then for the fid model. The parameter $\alpha$ shows a less clear
pattern with redshift, but it is generally lowest for the fid model.
Overall, the most important implication is that the gas fraction in
halos is highly sensitive to the assumed initial conditions.

Comparing to linear theory allows us to understand some of these
results. As noted in section~\ref{sec:Mf}, we calculated the filtering
mass from linear theory for each of the ICs, and the linear
calculation allows us to understand the relative importance of
pressure in the various IC models, at least during the linear
evolution. Although the simulation results come from non-linear,
viralized halos, we find an approximate agreement (typically to within
$\sim 20\%$) between the filtering mass, as defined here and in our
previous work \citep{NBM,NB07}, and the characteristic mass as
measured in the simulation, for all the models. In particular, the
relative sizes of $M_c$ among the various models, and the slow decline
of all the characteristic masses with time, are well matched by the
corresponding $M_F$ values predicted from linear theory. This close
match can be understood from the fact that while both gravity and
pressure increase during the non-linear evolution, their relative
strength only changes by a relatively small factor as a halo undergoes
non-linear collapse and virialization. Halos in which pressure had a
large effect during the early, linear evolution stage, keep sufficient
pressure to maintain the suppressed baryon content all through the
final collapse. On the other hand, in more massive halos in which
gravity overcame pressure early on, the baryons keep up with the
collapse of the dark matter and the pressure never has a major role.

For the E-$\delta$ alternative model, we find that the resulting
characteristic mass is higher than the result in the fid
model. Specifically, at $z=20$ we find $M_c\sim 5\times
10^4$~M$_\odot$ and $\alpha\sim 1$). This can be understood since
setting the gas fluctuations to be equal to the dark matter's means
that the pressure of the gas is higher compared to the fid model. As
can also be seen from comparison to linear theory, the system retains
the memory of the pressure, due to the time integrated nature of the
filtering mass. Therefore, the higher pressure translates to a higher
filtering/characteristic mass.

The mean $c_s$ approximation starts with effectively smoother ICs than
in the fid model ($\sim 20\%$ underestimate of the small-scale baryon
overdensity). Thus, the baryonic components lag behind the dark matter
collapse, and the pressure is always overestimated for a given baryon
overdensity (due to the overestimated temperature fluctuations),
resulting in a lower gas fraction for any given halo mass, i.e., the
characteristic mass is higher than in the fid model. Specifically, at
$z=20$ we find $M_c\sim 7\times 10^4$~M$_\odot$ and $\alpha\sim 1$.
This can be compared with $M_c\sim 3\times 10^4$~M$_\odot$ and
$\alpha\sim 0.6$ for the fid ICs.

Recently, \citet{Hoeft} and \citet{Okamoto} showed that the
characteristic mass scale does not agree with the \citet{cs} filtering
mass in the low-redshift, post-reionization regime. However, it is
important to note that at these low redshifts, the heating/cooling and
other feedback mechanisms are complex and highly inhomogeneous, so
that the ``filtering mass'' calculated from linear theory is not
really precisely defined, and the comparison of the linear and
non-linear results cannot really be considered a direct and precise
test. In contrast, \citet{NBM} found that the filtering mass gives a
good approximation to the characteristic mass, even in the presence of
a "flash" heating event \citep[see also][]{Andrei} that is physically
somewhat contrived but allows for a clear comparison of the linear and
non-linear results.

Summarizing our results, we find a good agreement between the
characteristic mass and the filtering mass in all the models.  Figure
\ref{fig:Mc} shows the best fitted parameters at various redshifts for
$M_c$ and $\alpha$, and our value for $f_{\br,0}$, for all models (the
1-$\sigma$ ($68\%$) confidence regions are listed in
table~\ref{table1}). It is important to emphasize that in this
statement we are referring to our definition in equation~(\ref{Mf}),
which is one eighth of the original definition which \citet{gnedin00}
claimed was a good fit to the characteristic mass. While we have been
careful to select halos with at least 500 particles, based on the
results of \citet{NBM}, we do not have the even higher mass resolution
needed to perform a resolution convergence test as they did. Our main
conclusion is that at least in the redshift range $z=11-21$ the
filtering mass provides a fairly good estimate for the characteristic
mass. This extends the redshift range of the agreement between the
filtering mass and the characteristic mass found in
\citet{NBM} ($z=20-25$). Another significant result from this
agreement is that previous work (either analytical, semi-analytical,
or using simulations) that used the filtering or characteristic mass
without accounting for the correct initial conditions resulted in
inaccurate results. This is due to the significant (factor of 2--3)
variation among the predictions of the filtering/characteristic mass
in the various models. Since this mass scale is of prime importance in
early structure formation it is imperative to calculate it accurately.

\section{Discussion}\label{dis}

We have used three-dimensional hydrodynamical simulations 
to investigate the effect of different initial conditions
on the gas fraction in halos in the early universe. 
Specifically, we studied the minimum ``gas-rich'' mass 
defined to have half of the mean cosmic baryon fraction. 
We tested three different models for the initial
conditions  (see text for more details) \begin{enumerate}
    \item "fid"  ICs; this model is based on the linear
evolution from \citet{NB05}, which allows the baryonic 
speed of sound to spatially vary as a
result of the Compton scattering with the CMB.
    \item "E-$\delta$" ICs; in this model, the linear evolution
    from \citet{NB05} is modified to match a common assumption in
    the literature, where the linear initial overdensity of the
    baryons is taken to be equal to that of the DM, i.e., 
$\delta_\br=\delta_\dm=\delta_{tot}$, while conserving $\sigma_8$ from the fid model.
    \item "mean $c_s$ ICs"; this model assumes that the baryonic speed of sound is spatially
    uniform. Although \citet{NB05} showed that this assumption yields an inaccurate 
evolution of the baryon density and temperature perturbations, it is still often 
used in codes that generate initial conditions for simulations.
\end{enumerate}
For all of the tests we used a total of $2\times768^3$ particles of
dark matter and baryons with a box size of $2$~Mpc, starting at $z=99$.

There are two major findings from the analysis we present here. The
first, shown throughout the paper, is the importance of assuming the
correct initial conditions, both for analytical calculations and
numerical simulations.  Structure formation (both in the linear and
non-linear regime) and halo gas fractions are very sensitive to the
initial conditions even at relatively low redshifts ($\sim 10$). The
second major finding is the apparent agreement between the filtering
mass and the characteristic mass (to within $\sim 20\%$). This
suggests, as a broader implication, that one can use linear theory in
order to predict the overall trend of highly non-linear behavior (at
least in the case of determining the gas fraction of halos).

The the fiducial calculation, which was presented in
\citet{NB05}, follows the time evolution of the linear overdensities
correctly. However, the other ICs produce different results for the
baryonic structure formation. For instance, the non-linear power
spectrum (fig.~\ref{fig:nonlinear}) shows that the system still
remembers its initial condition differences even at redshift 15.  In
particular, the $c_s$ model underestimates the non-linear baryonic
fluctuations by about $10 \%$ while the E-$\delta$ model overestimates
them by $40\%$ on small scales.

The mean $c_s$ approximation and the E-$\delta$ model are often used
to set the initial conditions in simulations, e.g., the CMBFAST code
\citep{cmbf} assumes the mean $c_s$ approximation while
\citet{EH99} is used with the E-$\delta$ assumption.  
We have shown that the non-linear evolution is very sensitive to the
initial conditions (figure~\ref{fig:nonlinear}) and they affect the
gas fraction in small halos down to redshift $\sim 10$
(figure~\ref{fig:Mc}). Our results emphasize the importance of the
differences between the dark matter and baryons and of the spatial
sound speed fluctuations, in both the linear calculation and the
initial conditions of the simulations.

It is important to emphasize that although Compton heating is not
included in the Gadget code that we used in this analysis (Gadget-2),
the fiducial calculation still describes fairly well the
non-linear behavior. Actually, the Compton heating contribution to the
heating of the gas in non-linear objects is negligible compare to the
adiabatic heating due to the gravitational collapse (see Appendix
\ref{Cheating}). Also, as noted above, much of the contribution to
the filtering mass comes from the highest redshifts, above our
simulation starting redshift of 99, since the Jeans mass is highest
then and so the pressure has the greatest impact at that early time.

In each simulation, we calculated the characteristic mass for which a
halo keeps most of its baryons (eq.~\ref{f_g-alpha}). We found that
the fid calculation gives the lowest value, which suggests that with
these correct ICs, pressure plays only a moderate role in galaxy
formation. In particular, the characteristic mass of $\sim 3 \times
10^4$~M$_\odot$ is significantly below the minimum mass for molecular
hydrogen cooling, so the gas content is not strongly suppressed even
in the smallest star-forming halos. In other words we find that before significant
heating took place the baryon fraction in halos is (eq.~\ref{f_g-alpha} with $M_c\sim3\times10^4$~M$_{\odot}$ and $\alpha\sim 0.64$)
\begin{equation}
M_{\br}\sim M_{\rm tot} f_{\br,0} \bigg[1+0.16\left(\frac{3\times 10^4 M_{\odot}}{M}\right)^{16/25} \bigg]^{-75/16} \ .
\end{equation}
The other alternative models give
incorrect higher value for the characteristic mass, closer to the
minimum mass for forming stars. Even with the fid ICs, pressure does
strongly limit the amount of gas in minihalos below the molecular
hydrogen cooling mass. We note that this value of $3 \times
10^4$~M$_\odot$ assumes adiabatic evolution, in particular with no
stellar heating. This value is consistent with the results of
\citet{NBM} for a somewhat higher redshift range.

We find that the theoretical linear filtering mass (as defined in
section~\ref{sec:Mf}) is in fairly good agreement with the
characteristic mass. This finding is true for all the models tested
here, throughout a significant redshift range, so this may imply more
generally a close relation between linear theory and non-linear halo
formation. In addition, this is consistent with the findings by
\citet{NBM} from AMR simulations, where the
filtering mass and the characteristic mass agreed in the "E-$\delta$"
model, even when a sudden heating was introduced.


Finally, we emphasize that our results are valid only in the
pre-reionization era. At the end of the reionization, \citet{MD08}
concluded that the characteristic mass is likely to be close to the atomic-cooling
threshold of $\sim 10^8 M_{\odot}$, which is also close to the values
found by \citet{Hoeft} and \citet{Okamoto}.  

%

Recently \citet{Hirata} argued that the initial velocity difference
between the baryons and dark matter after recombination has not been
fully accounted for, because of a higher-order contribution that is
not included in the linear theory approach. They estimated this
higher-order effect within the mean $c_s$ approximation and found that
it causes an additional suppression of the small-scale power spectrum,
in turn affecting the formation of the first structures. This effect
should be further investigated as in our detailed approach here,
although this would be more difficult (analytically, it is a
higher-order and anisotropic term, and to simulate it directly would
require starting at quite high redshifts).

\section*{Acknowledgments}
We thank the anonymous referee for useful and helpful comments.
We thank Ikkoh Shimizo for supplying the code for calculating the
non-linear power spectrum. We also thank Nick Gnedin, Andrey Kravtsov, Matt McQuinn and  Michele Trenti  for useful discussions. SN also expresses special
thanks to Yoram Lithwick for interesting discussions.  The authors
acknowledge financial support by the Grants-in-Aid for Young
Scientists (S) 20674003 by the Japan Society for the Promotion of
Science.  SN acknowledges NASA ATP grant NNX07AH22G and in part the German-Israeli 
Project Cooperation (DIP) grant STE1869/1-1.GE625/15-1 and the generous
support of the National Post Doctoral Award Program for Advancing
Women in Science (Weizmann Institute of Science). 
RB is grateful for
the support of Israel Science Foundation grant 823/09.

\appendix
\section{$\sigma_8$ conservation}\label{sig8}

\begin{figure}
  \centering \includegraphics[width=84mm]{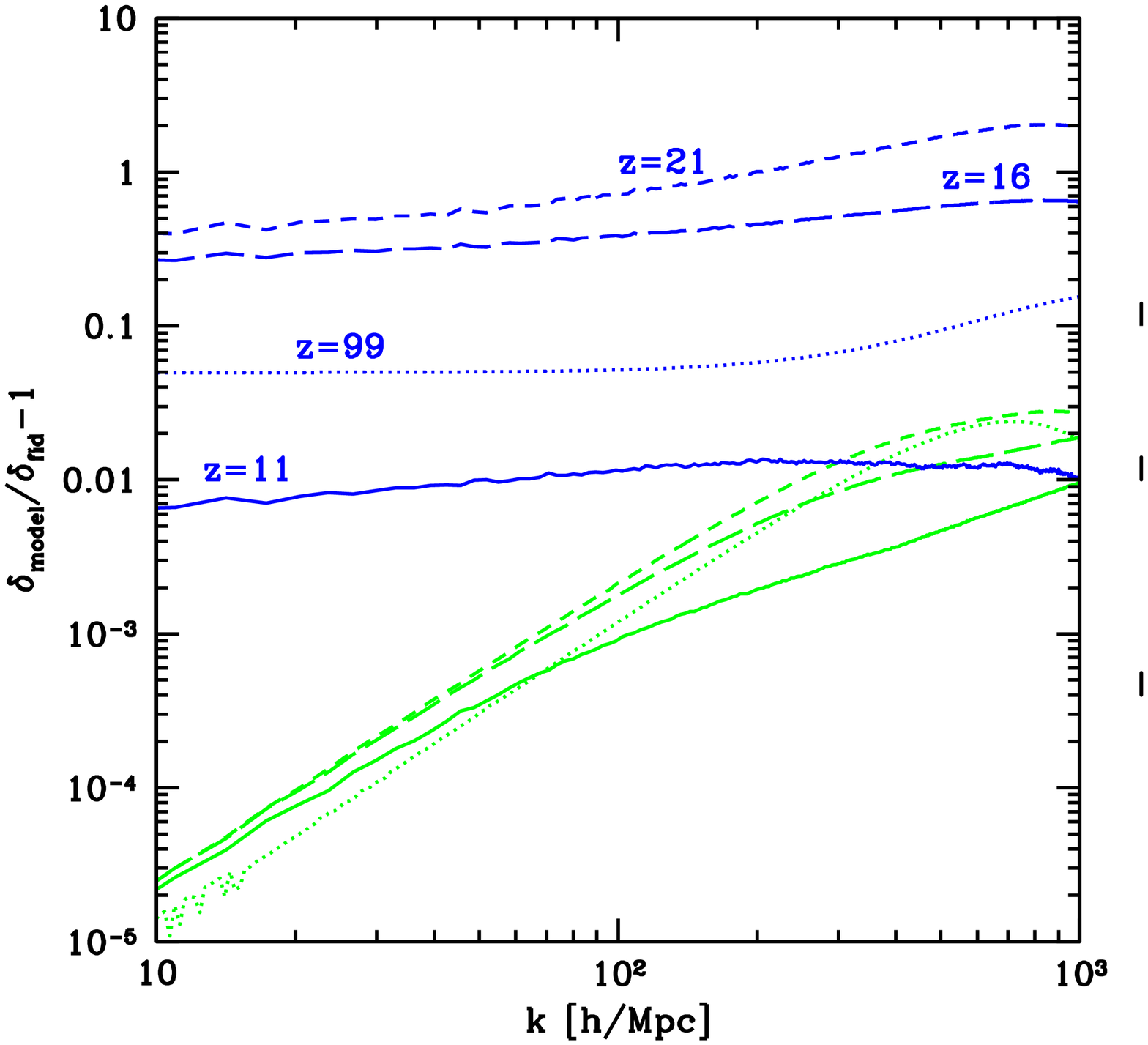}
\caption{ Demonstration of $\sigma_8$ conservation. The ratio of the {\em total} non-linear fluctuation for different redshifts (specifically $\delta_{\tot,model}/\delta_{\tot,{\rm fid}} -1$). We show the E-$\delta$ (upper set of $4$ curves) and the mean $c_s$ model (lower set of $4$ thin curves). We consider redshift $z=99, 21,16$ and $11$; dotted, short dashed, long dashed and solid curves respectively (labeled for the E-$\delta$ model).   }
\label{fig:PSev}
\end{figure}

We have defined the two different models such that they conserve
$\sigma_8(z = 0)$. From linear theory we do not expect the evolution
of the mean $c_s$ model to be significantly different from that of the
fid model (in terms of halo abundance and total power spectrum). This
is indeed the case for the evolution in time of the total fluctuations
of the mean $c_s$ model compared to the fid model on large scales (small
$k$), as shown in figure  \ref{fig:PSev} (lower set of thin curves).
%
%
%

A more delicate treatment is needed for the E-$\delta$ model (see
section \ref{sec:Edelta}). In this case, at high redshift (such as the initial $z =
99$), the baryons are in the process of falling into the DM
potential. This results in a faster growth of the total fluctuations
compared to the case in which there is a relative velocity between the 
DM and the baryons (such as in the case of the mean cs and fid models, where the relative velocity for the E-$\delta$ model are negligible); see figure \ref{fig:PSev}
dotted thick curve. At later times, the baryon fluctuations approach
the dark matter fluctuations, and the large scale behavior (i.e., on
linear scales) deviates from the fid model by less then $0.7\%$ (see the
solid curve in fig.~\ref{fig:PSev}).

 
We also note that we have checked the overall effect of $\sigma_8(z =
0)$ on the main results. We have performed two additional simulations
for the E-$\delta$ model, where we increased or decreased $\sigma_8$
by $5\%$. We found that the calculated $M_c$ is within the fit errors
(see appendix \ref{app} and table \ref{table1}) at $z > 12$. At $z \le 12$, the
difference in the best fitted value is below $0.5\%$."

%
%

\section{ Fit properties}\label{app}
\begin{table}
 \caption{ The best-fit parameters from equation~(\ref{f_g-alpha}).}
\label{table1}
\begin{center}
\begin{tabular}{l c c c }
\hline
Redshift & $M_c$   [$10^4$~M$_\odot$]              &  $\alpha$ \\
\\
\hline \hline
&  {\bf fiducial }  &\\
&  {\bf calculation}  &\\
\hline \hline
$21$ & $5.7^{+9.9}_{-5.3}$&  $0.7^{+0.45}_{-0.45}$\\
$20$ & $3.2^{+5.6}_{-1.9}$&  $0.61^{+0.39}_{-0.39}$ \\
$19$   & $2.6^{+5.4}_{-2.1}$&  $0.77^{+0.39}_{-0.5}$ \\
$18$ & $2.8^{+2.3}_{-2.4}$&  $0.77^{+0.03}_{-0.23}$\\
$17$   & $3.7^{+1.6}_{-1.6}$ &  $0.84^{+0.02}_{-0.24}$\\
$16$   & $4.4^{+2.4}_{-2.2}$&  $0.77^{+0.21}_{-0.14}$  \\
$15$   & $3.2^{+1.3}_{-1.3}$&  $0.7^{+0.1}_{-0.24}$  \\
$14$   & $3.4^{+0.4}_{-1.2}$&  $0.75^{+0.12}_{-0.09}$  \\
$12$   & $2.9^{+0.2}_{-0.2}$ &  $0.69^{+0.1}_{-0.15}$\\
$11$   & $3.2^{+0.1}_{-0.1}$ &  $0.78^{+0.1}_{-0.14}$\\
\hline \hline
&  {\bf mean }  {\bf $c_s$} &\\
\hline \hline
$21$ & $13.4^{+10.7}_{-7.5}$&  $1.23^{+0.52}_{-0.72}$ \\
$20$ & $7.2^{+5}_{-5.5}$&  $1.18^{+0.06}_{-0.92}$\\
$19$   & $9.4^{+5.5}_{-4.9}$&  $1.07^{+0.32}_{-0.9}$\\
$18$   & $8.9^{+5.5}_{-6}$&  $0.98^{+0.62}_{-0.31}$\\
$17$   & $8^{+3.4}_{-3.3}$ &  $0.92^{+0.3}_{-0.2}$\\
$16$   & $6.6^{+3}_{-2.6}$&  $0.69^{+0.26}_{-0.53}$\\
$15$   & $5.9^{+2.1}_{-2}$&  $0.69^{+0.1}_{-0.26}$\\
$14$& $7.3^{+1.6}_{-1.6}$&$0.94^{+0.12}_{-0.09}$\\ 
$12$& $4.9^{+0.8}_{-0.1}$&$0.74^{+0.06}_{-0.04}$\\ 
$11$& $5.7^{+0.7}_{-0.7}$&$0.86^{+0.06}_{-0.03}$\\ 
\hline \hline
&  {\bf E-$\delta$ } &\\
\hline \hline
$21$ & $6.3^{+12.2}_{-4.8}$&  $1.04^{+1.5}_{-0.56}$ \\
$20$ & $5^{+9.6}_{-4.8}$&  $1.03^{+1.4}_{-0.51}$\\
$19$   & $6.2^{+5}_{-4.8}$&  $1.08^{+1.2}_{-0.45}$\\
$18$   & $5.5^{+4.5}_{-4.8}$&  $1.01^{+1}_{-0.3}$\\
$17$   & $4.9^{+3.2}_{-3.1}$ &  $0.92^{+0.42}_{-0.21}$\\
$16$   & $4.4^{+2.5}_{-2.4}$&  $0.93^{+0.31}_{-0.18}$\\
$15$   & $4.2^{+1.7}_{-1.7}$&  $0.89^{+0.19}_{-0.12}$\\
$14$& $5.4^{+1.5}_{-1.7}$&$1.11^{+0.13}_{-0.19}$\\ 
$12$& $4.9^{+0.8}_{-0.7}$&$1.06^{+0.06}_{-0.06}$\\ 
$11$& $5.3^{+2.2}_{-0.8}$&$1.11^{+0.05}_{-0.05}$\\ 
\hline \hline
\vspace{-0.7cm}
\end{tabular}
\end{center}
\end{table}

\begin{figure}
  \centering \includegraphics[width=84mm]{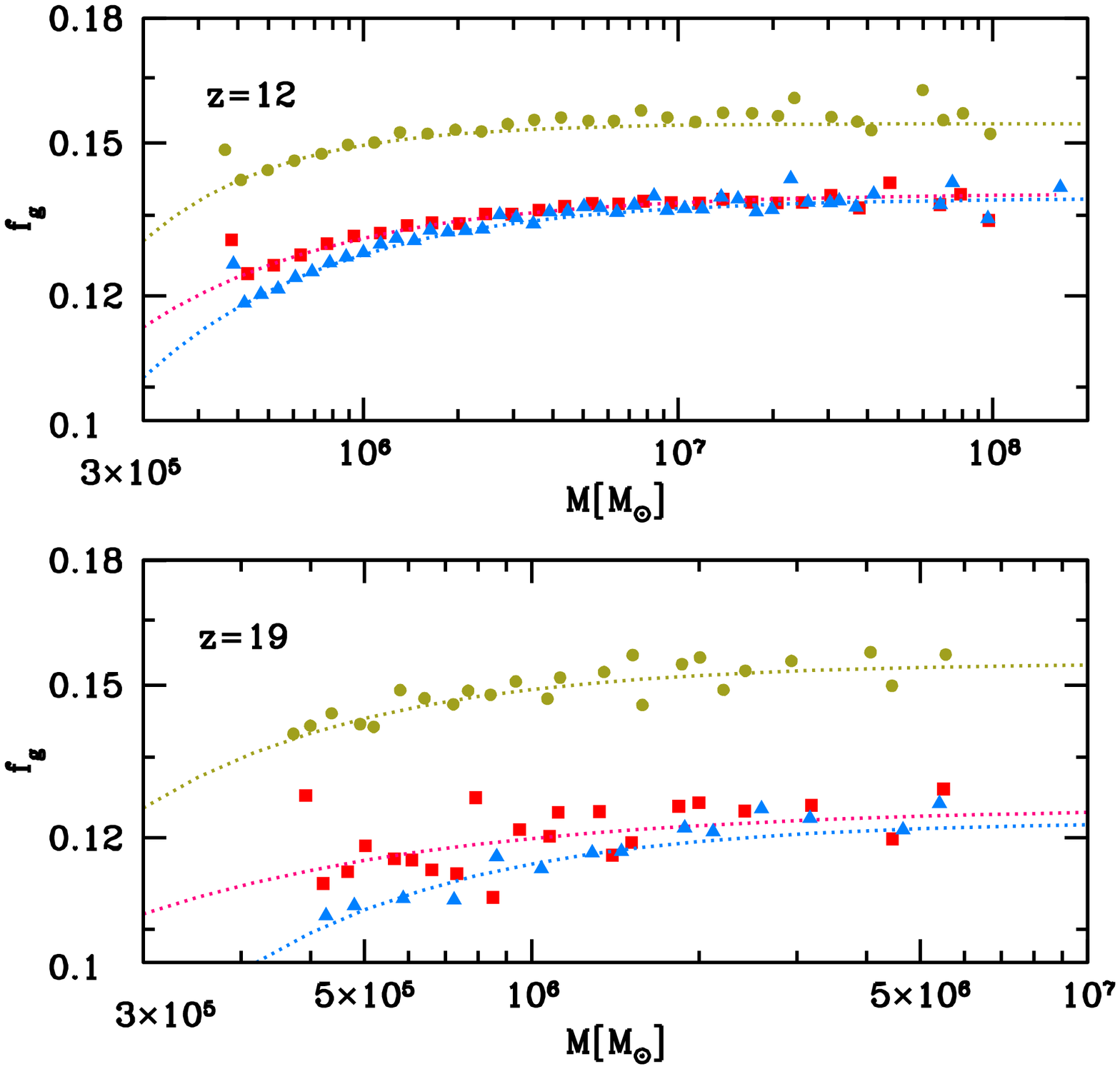}
\caption{ Two redshift examples of fitting the characteristic mass
($z=19$ and $z=12$). We consider the fiducial calculation,
mean $c_s$ approximation and the E-$\delta$ model (boxes, triangles
and circles, respectively), where we fit equation~\ref{f_g-alpha} to
all data points from halos with at least 500 particles. We also show
the fits from table~\ref{table1} (dotted curves).  }
\label{fig:fit}
\end{figure}

For each redshift snapshot for each run we find the characteristic
mass and $\alpha$ using a two dimensional fit. In figure \ref{fig:fit}
we consider two example redshifts (high, $z=19$ and low, $z=12$) for
which we show the binned data points and the resulting fit. In table
\ref{table1} we show our best fit parameters.  We note that we have 
checked that the fits give consistent results if we lower the
condition on the minimum number of particles per halo to 300 (instead
of 500). 
We also note that our determination of $M_c$ relies on an extrapolation (via the fit) below our
 simulations' resolution limit

The parameter $f_{b,0}$ in equation (\ref{f_g-alpha}) is an average of
the gas fraction values in the few highest mass bins. In our
simulation the high-end tail of the masses has large scatter in the
estimated gas fraction because of the low number of halos (each bin
among the last 3 or 4 in figure~\ref{fig:fit} represents just 1 or 2
halos), thus we have to average over this scatter to get a reasonable
result.  This scatter is in part a result of assuming that the halos
are spherical, and thus halos that are undergoing a major merger
deviate greatly from a spherical shape and are treated inaccurately in
our analysis. We have tested the resulting $f_{b,0}$ when taking a linking parameter
of $0.1$, which indeed resulted in more high-mass halos, but in any
case was consistent with the value of $f_{b,0}$ we found with the
$0.2$ linking parameter. Thus, in this paper, we use the standard
value of $0.2$.

As expected at high redshift, where we have fewer halos, the errors
become quite large. We also tried, following \citet{NBM}, to bin the
data and to perform the fit for the binned data with the $1-\sigma$
weight for each bin. For the redshifts for which we had more than
$\sim 1000$ halos we got that the binned analysis gave results within
the non-binned fit errors,and with comparable  errors.
 
We also tried the approach of taking $f_{\br,0}$ to be a free
parameter, but this produced very problematic
fits\footnote{\citet{NBM} also found that treating $f_{\br,0}$ as a
free parameter was unproductive. }.  This is mainly because of the
large scatter at the high mass end, so that a three-parameter fit
could not strongly constrain the parameter values. We also note the
fact that $f_{b,0}$ is lower than the mean cosmic fraction
$\bar{f}_b$, by about $20\%$ - $12\%$ for the fid and mean $c_s$
models, and $\sim 5\%$ for the E-$\delta$ model (see figure
\ref{fig:Mc} top panel). The result in the fid and mean $c_s$ 
models may reflect the real suppression of the large-scale baryon
fluctuations in these models; the difference in linear theory is $\sim
6\%$ at $z=20$ \citep{NB07}, but the non-linear evolution may increase
this effect. The discrepancy in the E-$\delta$ model may reflect a
limitation of the simulation; we note that in \citet{NBM} $f_{b,0}$
was also lower than $\bar{f}_b$ and even lower by $10\%$ from our results at the overlapping redshifts  (where we
compare the E-$\delta$ model in both cases). This might be due to the
fact that gas shocks in AMR are sharper than in Gadget simulations,
and thus AMR may produce a more realistic gas profile, although the
result is still below the universal cosmic baryon fraction
\citep{lin+06}. In our simulation, going to a larger radii can result
in a more realistic value, but we used $R_{200}$ for 
consistency with the common definition.

\section{Heating of non-linear halos}\label{Cheating}

The fiducial model follows correctly the baryon density and
temperature perturbations due to Compton scattering on the residual
free electrons after recombination. While this is fully incorporated
in our fid ICs, our simulation does not take into account Compton
heating. Below we show that for non-linear objects the heating is
actually negligible compared to the adiabatic heating due to the
gravitational collapse of baryons into the dark matter potential
wells. Therefore, it is sufficient to include Compton heating in
the linear stage only.

The heating of the gas $Q_{\mathrm comp}$ due to Compton heating from
the CMB \citep{NB05} during the free-fall time $1/\sqrt{G\rho}$ of
gravitational collapse is
\begin{equation}
\label{h_rate}  Q_{\mathrm comp} \propto 4 \frac{\sigma_{T}\, c} {m_e } \, 
k_B (T_\gamma - T) \rho_\gamma x_e(t)\frac{1}{\sqrt{G\rho}} \ ,
\end{equation}
where $\sigma_{T}$ is the Thomson scattering cross section,
$\rho_\gamma$ is the photon energy density, $T_\gamma$ and $T$ are the
CMB and gas temperature and $x_e(t)$ is the electron fraction out
of the total number density of gas particles at time $t$.

The virial theorem gives a relation in collapsed objects between the
thermal energy $E_{th}$ and the gravitational energy $E_{gr}$, i.e.,
$E_{th}=-E_{gr}/2$ . Thus, for a halo mass $M$ with virial radius
$r_{vir}$ the thermal energy can be expressed as:
\begin{equation}
E_{th}\sim \frac{1}{2}\frac{G M ^2}{r_{vir}} \ .
\end{equation}
For all relevant redshifts and mass scales we find that $Q_{\mathrm
comp}/E_{th} <<1$. Therefore, neglecting the contribution of the
Compton heating during the non-linear evolution is justified. However,
as we have shown, neglecting the Compton heating in the linear
evolution and in the initial conditions leads to inaccurate values for
the gas fraction in halos.

\end{document}